# The GALAH Survey: Stellar parameters and abundances for 800 000 *Gaia* RVS spectra using GALAH DR4 and *The Cannon*

Pradosh Barun Das [1,2,3]★ Daniel B. Zucker [1,2,3], Gayandhi M. De Silva [1,2,3], Nicholas W. Borsato [1,2,4], Aldo Mura-Guzmán [1,2,3], Sven Buder [3,5], Melissa Ness [3,5,6,7], Thomas Nordlander [3,5], Andrew R. Casey [3,8], Sarah L. Martell [3,9], Joss Bland-Hawthorn [3,10], Richard de Grijs [1,2,3,11], Ken C. Freeman [3,5], Janez Kos [12], Dennis Stello [3,9,10], Geraint F. Lewis [10], Michael R. Hayden [3,10,13] and Sanjib Sharma [14]

[1] *School of Mathematical and Physical Sciences, Macquarie University, Sydney, NSW 2109, Australia*
[2] *Astrophysics and Space Technologies Research Centre, Macquarie University, Sydney, NSW 2109, Australia.*
[3] *ARC Centre of Excellence for All Sky Astrophysics in 3 Dimensions (ASTRO-3D), Australia*
[4] *Division of Astrophysics, Department of Physics, Lund Observatory, Lund University, Box 118, SE 221 00 Lund, Sweden*
[5] *Research School of Astronomy and Astrophysics, The Australian National University, Canberra, ACT 2611, Australia*
[6] *Department of Astronomy, Columbia University, Pupin Physics Laboratories, New York, NY 10027, USA*
[7] *Center for Computational Astrophysics, Flatiron Institute, 162 Fifth Avenue, New York, NY 10010, USA*
[8] *School of Physics and Astronomy, Monash University, Melbourne, VIC 3800, Australia*
[9] *School of Physics, University of New South Wales, Sydney, NSW 2052, Australia*
[10] *Sydney Institute for Astronomy, School of Physics, A28, The University of Sydney, Sydney, NSW 2006, Australia*
[11] *International Space Science Institute – Beijing, 1 Nanertiao, Zhongguancun, Hai Dian District, Beijing 100190, China*
[12] *Faculty of Mathematics and Physics, University of Ljubljana, Jadranska 19, 1000 Ljubljana, Slovenia*
[13] *Homer L. Dodge Department of Physics and Astronomy, The University of Oklahoma, 440 W. Brooks St., Norman, OK 73019, USA*
[14] *Space Telescope Science Institute, 3700 San Martin Drive, Baltimore, MD 21218, USA*

**ABSTRACT**

Analysing stellar parameters and abundances from nearly one million *Gaia* Data Release 3 (DR3) Radial Velocity Spectrometer (RVS) spectra poses challenges due to the limited spectral coverage (restricted to the infrared Ca II triplet) and variable signal-to-noise ratios of the data. To address this, we use *The Cannon*, a data-driven method, to transfer stellar parameters and abundances from the GALAH (GALactic Archaeology with HERMES) DR4 ($R \sim 28\,000$) catalogue to the lower resolution *Gaia* DR3 RVS spectra ($R \sim 11\,500$). Our model, trained on 14 484 common targets, predicts parameters such as $T_{\rm eff}$, $\log g$, and [Fe/H], along with several other elements across approximately 800 000 *Gaia* RVS spectra. We utilize stars from open and globular clusters present in the *Gaia* RVS catalogue to validate our predicted mean [Fe/H] with high precision ($\sim 0.02-0.10$ dex). Additionally, we recover the bimodal distribution of [Ti/Fe] versus [Fe/H], reflecting the high and low $\alpha$-components of Milky Way disc stars, demonstrating *The Cannon*'s capability for accurate stellar abundance determination from medium-resolution *Gaia* RVS spectra. The methodologies and resultant catalogue presented in this work highlight the remarkable potential of the RVS data set, which by the end of the *Gaia* mission will comprise spectra of over 200 million stars.

## 1 INTRODUCTION

Galactic archaeology effectively bridges stellar spectroscopy – which allows for precise measurement of stellar parameters and chemical abundances – with studying the formation and evolution of the Milky Way galaxy (Freeman & Bland-Hawthorn 2002). Historically, high-quality elemental abundance analyses were typically restricted to at most a few thousand stars (Edvardsson et al. 1993; Bensby, Feltzing & Oey 2014). However, over the last decade, technological advancements and extensive spectroscopic surveys such as LAMOST (Large Sky Area Multi-Object Fiber Spectroscopic Telescope; Eisenstein et al. 2011), APOGEE (Apache Point Observatory Galactic Evolution Experiment; Majewski et al. 2017), *Gaia*-ESO (Gilmore et al. 2012), space-based *Gaia* (Gaia Collaboration 2016, 2018, 2023a) with its Radial Velocity Spectrometer (RVS, Recio-Blanco et al. 2023; Gaia Collaboration 2023b) and low-resolution XP spectra (Andrae, Rix & Chandra 2023), RAVE (the RAdial Velocity Experiment; Steinmetz et al. 2006), ARGOS (Abundances and Radial velocity Galactic Origins Survey; Ness et al. 2012), and

★ E-mail: pbdrohan@gmail.com

GALAH (GALactic Archaeology with HERMES; De Silva et al. 2015; Martell et al. 2017; Buder et al. 2018, 2021) have yielded the chemical compositions for millions of stars, with this number continuing to grow. In addition to these existing projects, multiple new extensive surveys such as SDSS-V (Sloan Digital Sky Survey; Kollmeier et al. 2017), 4MOST (4-metre Multi-Object Spectroscopic Telescope; de Jong et al. 2019), and WEAVE (William Herschel Telescope Enhanced Area Velocity Explorer; Dalton et al. 2018) will enormously expand the data available, allowing us to study the Milky Way in much greater detail. The detailed chemical abundances these surveys produce, provide unique chemical signatures of stars, offering direct insights into their birth environments (Krumholz, McKee & Bland-Hawthorn 2019) and potential chemical enrichment pathways (Rybizki, Just & Rix 2017). Such progress significantly enhances efforts to carry out chemical tagging – that is, identifying stars that are formed in the same (or similar) environment based on their abundance patterns (De Silva et al. 2015; Hawkins et al. 2015; Hogg et al. 2016; Buder et al. 2022), and facilitates more comprehensive analyses of stellar populations.

The vast quantities of data generated by these large spectroscopic stellar surveys however pose significant challenges in data analysis and modelling. Ranging from several million stars in LAMOST (Cui et al. 2012; Deng et al. 2012; Zhao et al. 2012) to approximately a quarter-billion stars observed in *Gaia* Data Release 3 (DR3) Blue Photometer/Red Photometer (BP/RP) spectra (Andrae et al. 2023; Zhang, Green & Rix 2023; Yao et al. 2024), many of these data sets are collected at low to moderate spectroscopic resolutions. This highlights the need for high-resolution spectroscopic data – the 'gold standard' for measuring stellar properties – to more accurately determine stellar parameters and elemental abundances. Beyond observing at different spectroscopic resolutions, these surveys also observe different wavelength regions, utilize different targeting strategies, and employ different pipelines and data analysis techniques to derive the characteristics of each observed star.

In this work, we will use the term 'labels' to refer collectively to these stellar characteristics – specifically parameters such as $T_{\rm eff}$ and $\log g$, and stellar elemental abundances derived from the spectra. These stellar properties are conventionally determined by comparing observed data to a set of model spectra with known characteristics (stellar atmospheres, line lists, etc., customized to each survey), using optimization techniques tailored to the specific wavelength range of a given survey (Boeche et al. 2011; Bailer-Jones et al. 2013; Mészáros et al. 2013; Liu et al. 2014). They typically focus on specific segments of the spectrum, prioritizing those absorption lines deemed the most reliable or relevant. After the initial analysis, additional procedures are often employed to refine the derived stellar properties using more dependable external information. Even when consistent initial assumptions are used, different analysis methods can still produce significant variations in derived stellar labels (e.g. Smiljanic et al. 2014). Consequently, many surveys use benchmark stars as references to assess the validity of their findings. However, significant discrepancies in the calibration of stellar labels can arise between different surveys or analysis pipelines, leading to differences in the scales of stellar attributes. These discrepancies complicate comparisons between surveys and present a significant challenge to researchers working in this era of large astronomical data sets.

The accuracy of these models can also be limited by the use of simplified physical assumptions. Some models may neglect important molecular opacities, stellar chromospheric effects, etc., resulting in discrepancies in the reported properties for the same stars using different wavelength regions, input assumptions, and methodologies. Although combining the results from different surveys has yielded significant scientific advancements, the disagreements between these surveys – in many cases because different pipelines measure substantially different labels for the same observed stars – pose a challenge for drawing conclusions based on heterogeneous stellar datasets (Ho et al. 2017; Nandakumar et al. 2022; Thomas et al. 2024).

Recently, velocities from nearly 33 million RVS spectra and over 200 million BP/RP spectra were published as part of *Gaia* DR3. The processing of the RVS spectra and the various derived data products are detailed in a series of papers: Katz et al. (2023) discuss the properties and validation of the radial velocities; Blomme et al. (2023) focus on the radial velocities of hot stars; Frémat et al. (2023) describe the properties of the broadening velocity $v_{\rm broad}$ derived with the RVS; and Sartoretti et al. (2023) explain the determination of $G_{\rm RVS}$ magnitudes. From these 33 million RVS spectra, approximately 1 million normalized RVS spectra were released, which were initially analysed by the 'General Stellar Parametriser-Spectroscopy' (GSP-Spec; Recio-Blanco et al. 2016, 2023). However, the GSP-Spec module did not yield accurate stellar abundances ([$\alpha/M$]) or atmospheric parameters for nearly one-third of the sample with low signal-to-noise ratios (S/N), i.e. within the range $15 \leq \rm{S/N} < 25$ (Guiglion et al. 2024); detailed information on the quality flag chain 'flags_gspspec' implemented for the stellar labels for the GSP-Spec pipeline is provided in table 2 of Recio-Blanco et al. (2023). In this context, less traditional methods – e.g. machine learning – can offer significant advantages for spectroscopic analysis of large survey data sets, enabling simultaneous prediction of stellar labels within a multidimensional label space. Some notable recent examples of such approaches include *The Payne* (Ting et al. 2019), which uses a training model incorporating an Artificial Neural Network (ANN) Interpolator (Bailer-Jones, Irwin & von Hippel 1998); the hybrid Convolutional Neural Network (CNN) method (Guiglion et al. 2024); and *The Cannon* (Ness et al. 2015; Casey et al. 2016; Ho et al. 2017), which utilizes a generative training model to link stellar spectra with the stellar labels.

For our investigation, we utilize the *The Cannon* to derive stellar properties from spectra, particularly in this context of massive spectroscopic surveys. This approach offers several notable advantages, including minimizing the direct reliance on physical models of spectra being analysed, computational efficiency, capacity to deliver precise stellar properties even at lower S/N, and its adeptness at harmonizing and calibrating disparate surveys to yield consistent outcomes (Ness et al. 2015). The effectiveness of *The Cannon* relies on the inclusion of 'reference objects' within the survey, for which stellar labels are already known, preferably from another survey with higher S/N values and/or higher spectral resolution. These labels provide essential insights into the characteristic features of stellar spectra and contain the trends and relationships to which *The Cannon* algorithm can, for example, map the relatively low-quality *Gaia* RVS spectra (typically low S/N, and low resolution, ∼ 11 500). Additionally, *The Cannon* assumes that objects with identical labels have similar spectra, exhibiting smooth variations with changes in the stellar labels. It largely depends on the selection of the training sample for the predicted stellar labels. Therefore, to ensure a wider coverage of labels, the training sample is expected to include different populations of stars with reliable and high-quality stellar characteristics from the reference surveys to ensure a broader representative sample of individual stars.

*The Cannon* has previously been applied to combine multiple large surveys. Ho et al. (2017) discuss the transfer of stellar labels from APOGEE (high resolution: $R \sim 22\,500$) to LAMOST (low resolution: $R \sim 1800$) spectra, such as $T_{\rm eff}$, $\log g$, [Fe/H], and [$\alpha/M$] for nearly 450 000 giants using *The Cannon*. The masses and ages for 23 000 giants (from LAMOST) were obtained using reference stars common to both APOGEE and LAMOST and employing stellar labels, including carbon and nitrogen abundances, with high precision and accuracy. Nandakumar et al. (2022) applied *The Cannon* to construct a combined database consisting of both APOGEE and GALAH-scaled stellar parameters. Using *The Cannon*, Wheeler et al. (2020) estimated the abundances for nearly 3.9 million LAMOST stellar spectra across five nucleosynthetic channels using elemental abundances from GALAH as reference data. Ness et al. (2016), Buder et al. (2018), and Hasselquist et al. (2020) also used *The Cannon* to derive precise stellar parameters and abundances between multiple surveys, using training samples taken from stars present in both reference and target surveys.

In this study, we utilize *The Cannon* 2 (Ness et al. 2015; Casey et al. 2016) to obtain stellar parameters ($T_{\rm eff}$, $\log g$, and $v \sin i$) and elemental abundances ([Fe/H], [Ca/Fe], [Si/Fe], [Ni/Fe], and [Ti/Fe]) for 796 633 *Gaia* RVS spectra using the high-quality stellar labels from GALAH DR4 (Buder et al. 2024) by employing a training sample of 14 484 stars common to both surveys.

The structure of the paper is as follows: in Section 2, we present the data used for our model, *Gaia* RVS (DR3) spectra, and GALAH DR4 labels. Section 3 provides an overview of *The Cannon*'s methodology, including details on the training sample selection, comparisons between the two survey results and details of various flagging options. In Section 4, we validate our *Cannon*-derived metallicities using members of a sample of open and globular clusters. Finally, in Section 5, we recover the bimodal distribution in [Ti/Fe] versus [Fe/H] observed in our Milky Way galaxy, demonstrating the different trends for the high- and low-$\alpha$ discs. This marks one of the first such detections using *Gaia* RVS spectra, and illustrates the potential for RVS data in exploring the Milky Way's stellar populations.

## 2 DATA

The main data catalogue analysed in this study consists of $\sim 1$ million RVS stellar spectra released as part of *Gaia* DR3. For our application of *The Cannon*, we have selected a set of stellar parameters ($T_{\rm eff}$, $\log g$, and $v \sin i$) and chemical abundances ([Fe/H], [Ca/Fe], [Si/Fe], [Ni/Fe], and [Ti/Fe]) from GALAH DR4.

### 2.1 *Gaia* RVS

The *Gaia* RVS instrument (Cropper et al. 2018; Gaia Collaboration 2018, 2023a) measures stellar radial velocities with high precision achieving accuracies of a few km s$^{-1}$ for stars down to 17th magnitude. Along with *Gaia*'s astrometric and photometric measurements, these accurate radial velocities enable comprehensive 3D kinematic studies of stars. The RVS spectra can also provide valuable information on stellar parameters and chemical compositions (e.g. Recio-Blanco et al. 2023).

In this work, we analysed 999 645 normalized and radial velocity-corrected *Gaia* RVS spectra[1] (averaged over multiple transits; Seabroke et al. 2021) from *Gaia* DR3 (Katz et al. 2023), as well as their flux uncertainties per wavelength pixel. Cropper et al. (2018) provide a historical overview of the *Gaia* RVS. Each spectrum (spectral resolving power of approximately $R \sim 11\,500$; Cropper et al. 2018) contains 2401 pixels per scan with a pixel size of 0.10 Å, covering a total spectral range of 240 Å from 8460 to 8700 Å. We excluded potential galaxies and quasars using the flags `in_galaxy_candidates = False`, and `in_qso_candidates = False`. Additionally, we filtered out objects showing variability (`phot_variable_flag ≠ VARIABLE`) and signs of binarity (`non_single_star = 0`). We also removed stars with negative parallaxes and high photometric errors. After applying the above-mentioned conditions, our final sample included 796 633 RVS spectra for our analysis. Fig. 1(a) shows the S/N distribution for the 796 633 *Gaia* RVS targets, revealing that a significant fraction (nearly one-third) of the total sample has low spectral S/N, between 15 and 25.

[1] https://doi.org/10.17876/gaia/dr.3/54

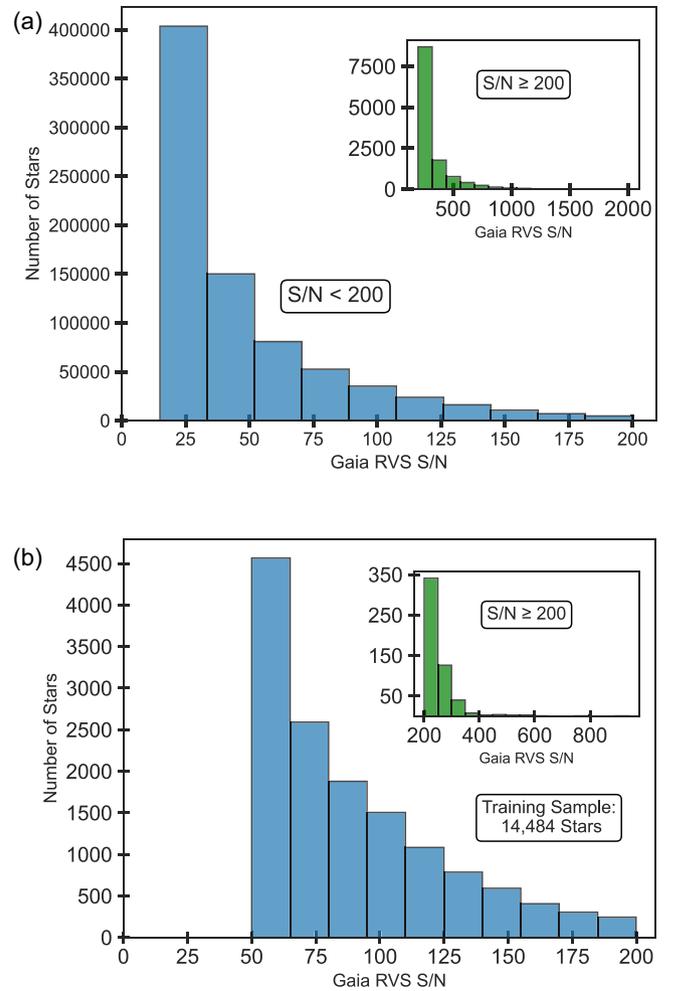

**Figure 1.** S/N distribution for the *Gaia* RVS spectra. (a) S/N distribution for our selected sample of 796 633 *Gaia* RVS spectra. The primary plot shows the distribution of the targets having S/N < 200 pixel$^{-1}$ element. The subplot at the top-right corner represents the distribution of the targets with S/N ≥ 200. We observe that one-third of the total sample of 796 633 RVS target spectra was observed with low S/N (between 15 and 25). (b) S/N distribution of the training sample of 14 484 stars used for *The Cannon* with S/N > 50.

### 2.2 GALAH Data Release 4

The GALAH survey (De Silva et al. 2015; Martell et al. 2017; Buder et al. 2018, 2021) is a comprehensive high-resolution spectroscopic survey conducted with the High Efficiency and Resolution Multi-Element Spectrograph (HERMES; Brzeski, Case & Gers 2011; Heijmans et al. 2012; Sheinis et al. 2015) on the 3.9 m Anglo-Australian Telescope. HERMES delivers high-resolution ($R \sim 28\,000$) spectra across four passbands for up to 400 stars simultaneously, utilizing a fibre positioner with a two degree field of view (2dF; Lewis et al. 2002). It produces spectra with S/N ∼ 100 per resolution element in an hour integration for stars with $V = 14$ mag. The four HERMES channels cover nearly 1000 Å in total: the blue channel (4718 − 4903 Å), the green channel (5649 − 5873 Å), the red channel (6481 − 6739 Å), and the infrared channel (7590 − 7890 Å). The GALAH survey primarily observes stars within 4 kpc of the Sun, focusing on those at Galactic latitudes $|b| > 10°$. Few stars are observed near the Galactic plane, and those are mainly in the direction of the Galactic Centre (Buder et al. 2021).

We use Data Release 4 (DR4) of the GALAH Survey (Buder et al. 2024), which provides the chemical abundances for up to 31 elements for 827 288 stars. GALAH DR4 is a combination of GALAH Phase 1 (bright, main, and faint survey), GALAH Phase 2 (focusing on main-sequence turnoff stars), TESS–HERMES (Sharma et al. 2018), and K2-HERMES programs (Sharma et al. 2019), as well as a few selected observations of globular and open cluster members. For this work, we have chosen only those targets that satisfied the flag labels in GALAH DR4 recommended for scientific applications (`flag_sp = 0`, `flag_fe_h = 0` and `flag_x_fe = 0`, where $x$ denotes the element of interest for obtaining abundances relative to Fe).

## 3 METHOD

### 3.1 *The Cannon*

*The Cannon* (Ness et al. 2015; Ho et al. 2017) is a data-driven machine-learning algorithm that facilitates the development of a model for stellar spectra based on a shared set of training targets' spectra common to multiple surveys. This approach depends on having a subset of reference objects within a survey, for which the stellar labels are accurately known with high fidelity and which adequately span the required label space. It differs from other techniques – such as principal component analysis, $k$-nearest neighbours, and neural networks (including CNNs, etc.) – in the way that it explicitly incorporates a noise model into the process, which enables *The Cannon* to transfer stellar labels from high S/N training set stars to low S/N test set stars with accuracy; in other words, the training and test data need not be statistically identical. As a result, it provides a reliable method for survey cross-calibration and cross-validation by transferring stellar labels between the surveys.

During the training step, the spectral model coefficients are determined individually at each wavelength pixel while maintaining fixed labels for all star spectra within the training set. This results in a spectral model that characterizes the flux at each wavelength pixel as a function of the provided stellar labels. These labels encapsulate crucial insights into the characteristic features of the stellar spectrum, and their corresponding coefficient values signify the impact of the respective labels at a specific wavelength pixel. In the subsequent label inference phase, the label coefficients remain unchanged. Likelihood optimization is then performed to predict the labels based on the flux values observed at each wavelength pixel of every test spectrum. Using *The Cannon* 2[2] (Casey et al. 2016), we develop a quadratic model with the stellar labels, that constructs the spectral model in the form mentioned in equation (1). In this equation, $G_{n\lambda}$ denotes the flux at each wavelength pixel ($\lambda$) for each star ($n$) in the training sample, $f(l_n)$ represents the vectorizing function in the form of a quadratic polynomial, $\theta_\lambda$ represents the set of spectral model coefficients corresponding to the label combinations at each wavelength, and $l_n$ denotes the various stellar labels used to develop the model. The code obtains the normalized labels ($\hat{l}_n$) using equation (2) for the trained model function.

$$G_{n\lambda} = f(l_n).\theta_\lambda + \text{noise}, \quad (1)$$

$$\hat{l}_n = \frac{l_n - l_{n,50}}{l_{n,97.5} - l_{n,2.5}} \quad (2)$$

The noise in equation (1) is essentially the root-mean-square combination of two key components: the inherent uncertainty ($\sigma_{n\lambda}$) associated with each pixel's flux, arising from finite photon counts and instrumental effects, and the intrinsic scatter of the model at each wavelength ($s_\lambda$). It can be represented in the following form:

$$\text{noise} = [s_\lambda^2 + \sigma_n^2]\xi_{n\lambda}. \quad (3)$$

Here, for each spectrum at every wavelength pixel, $\xi_{n\lambda}$ denotes a Gaussian random number with zero mean and unit variance. The intrinsic scatter represents the expected deviation of the spectrum from the generative model at that pixel, even when the measurement uncertainty approaches zero. Other than flux variance at each $\lambda$ provided by the observed spectra, the excess variance is obtained together with $\theta_\lambda$ by optimizing the single-pixel log likelihood function, described in equation (4), for all the stars in the training sample.

$$\ln p(G_{n\lambda}|\theta_\lambda, f(l_n), s_\lambda^2)$$
$$= -\left[\frac{[G_{n\lambda} - f(l_n) \cdot \theta_\lambda]^2}{2(s_\lambda^2 + \sigma_{n\lambda}^2)} + \frac{\ln(s_\lambda^2 + \sigma_{n\lambda}^2)}{2}\right]. \quad (4)$$

Using $G_{n\lambda}$ and $f(l_n)$, the coefficients and the scatter of the spectral model can be obtained by optimizing equation (5).

$$\theta_\lambda, s_\lambda \leftarrow \underset{\theta_\lambda, s_\lambda}{\arg\max}\left[\sum_{n=1}^{N} \ln p(G_{n\lambda}|\theta_\lambda, f(l_n), s_\lambda^2)\right]. \quad (5)$$

Once the model is trained, *The Cannon* then predicts the stellar labels ($l_m$) corresponding to each test spectrum ($m$), by optimizing the following likelihood function throughout the wavelength region:

$$l_m \leftarrow \underset{l}{\arg\min}\left[\sum_{\lambda=0}^{N_{\text{pix}}} \frac{[G_{m\lambda} - f(l_m) \cdot \theta_\lambda]^2}{(s_\lambda^2 + \sigma_{m\lambda}^2)}\right]. \quad (6)$$

### 3.2 Training data set

We assembled a training data set comprising targets common to both the *Gaia* RVS and GALAH DR4 catalogues. Through cross-matching the `source_id` column from *Gaia* RVS and the `gaiadr3_source_id` column from GALAH DR4, we identified 14 484 stars by adhering to the criteria given below, to ensure high-quality stellar parameters and abundances.

We selected stars from the *Gaia* RVS dataset based on S/N (`rvs_spec_sig_to_noise`) ≥ 50 (see Fig. 1b). Within this subset, stars were further screened to ensure GALAH S/N ≥ 50 across all four CCD arms, while also satisfying additional criteria: a chi-squared value (`chi2_sp` from GALAH) < 4, *Gaia* Re-normalized Unit Weight Error (`ruwe`) < 1.2, parallax (`parallax`) > 0, broadening velocity (`vsini` > 0 from GALAH), and compliance with a number of flags derived from the GALAH data analysis – specifically, all selected stars adhere to the criteria that parameters such as `flag_sp`, `flag_fe_h`, and `flag_x_fe` (with 'x' representing the corresponding element of interest) are equal to zero. Furthermore, the sample was refined to include only stars with effective temperature ($T_{\text{eff}}$) ≤ 7000 K, as discussed in Buder et al. (2018), which highlighted for the 14 484 stars in our training set (common to both GALAH DR4 and *Gaia* RVS) using the stellar labels from the GALAH DR4 catalogue.

---
[2] https://github.com/andycasey/AnniesLasso

**Table 1.** Quality cuts adopted for the errors in GALAH stellar labels used for selecting the training sample of 14 484 stars common to both the *Gaia* RVS and GALAH DR4 catalogues.

| Error in stellar label | Quality cuts for errors from GALAH DR4 |
| --- | --- |
| err_$T_{\text{eff}}$ | < 80 K |
| err_$\log g$ | < 0.15 dex |
| err_[Fe/H] | < 0.10 dex |
| err_[Ca/Fe] | < 0.10 dex |
| err_[Si/Fe] | < 0.10 dex |
| err_[Ni/Fe] | < 0.10 dex |
| err_[Ti/Fe] | < 0.10 dex |

**Table 2.** Biases and RMSE values of the 12-fold cross-validation test for the training sample of 14 484 *Gaia* RVS Spectra trained by *The Cannon* using stellar parameters and abundances from GALAH DR4.

| Stellar label | RMSE | Bias |
| --- | --- | --- |
| $T_{\text{eff}}$ (K) | 86.56 | 12.27 |
| $\log g$ (dex) | 0.15 | 0.02 |
| $v \sin i$ (km $s^{-1}$) | 3.21 | 0.65 |
| [Fe/H] | 0.07 | 0.00 |
| [Ca/Fe] | 0.10 | 0.01 |
| [Si/Fe] | 0.07 | 0.00 |
| [Ni/Fe] | 0.05 | 0.00 |
| [Ti/Fe] | 0.07 | 0.00 |

### 3.3 Training the sample using *The Cannon*

Using the training sample, we trained *The Cannon* 2 (Ness et al. 2015; Casey et al. 2016) with eight parameters ($T_{\text{eff}}$, $\log g$, $v \sin i$, [Fe/H], [Ca/Fe], [Si/Fe], [Ni/Fe], and [Ti/Fe]) using a second-order polynomial *The Cannon* model. For example, a model with two parameters, $T_{\text{eff}}$ and $\log g$, will have the second-order polynomial formulated as equation (7):

$$f(\theta) = \theta_0 + [\theta_1 \cdot T_{\text{eff}}] \\ + [\theta_2 \cdot (T_{\text{eff}})^2] + [\theta_3 \cdot \log g] \\ + [\theta_4 \cdot T_{\text{eff}} \cdot \log g] + [\theta_5 \cdot (\log g)^2] . \quad (7)$$

The $\theta_n$ ($n \in [0, 5]$) in equation (7) are the model coefficients corresponding to the different combinations of the stellar labels, as described in equation (5). Similarly, in our model with eight stellar labels, the model incorporates eight label coefficients each for the linear and the quadratic terms, 28 coefficients for the cross-term labels and 1 constant term.

The bottom panel of Fig. 2 shows the Kiel diagram obtained using *The Cannon*-predicted stellar labels of the training data set. The overall structure of the diagram is quite similar to the panel above it (with GALAH values), indicating that the training sample yielded reasonable results for $T_{\text{eff}}$ and $\log g$. This was further analysed using a 12-fold cross-validation test (see Section 3.4) to ensure a robust representative sample for the label predictions of all the stars in the *Gaia* RVS sample.

### 3.4 12-fold cross-validation of the predictions

We conducted a 12-fold validation test on *The Cannon* 2 predictions for the training sample to infer more information on the reliability of the predicted stellar labels. In this process, the 14 484 targets were split into 12 groups, by assigning each one a random integer between 0 and 11. We left out each group in turn (a sample test set), and trained a model on the remaining eleven groups. We then applied that model to infer new labels for the sample test set that was omitted. This process enabled us to get information on the root-mean square error (RMSE) values and the corresponding biases (residual of the average estimated value compared with the values from GALAH) in the model developed for the training sample (see Table 2). Fig. 3 shows the one-to-one plots of *The Cannon*-predicted stellar labels versus input GALAH DR4 stellar labels, and the lower plot in each of the panel for the stellar labels shows the distribution of the residuals,

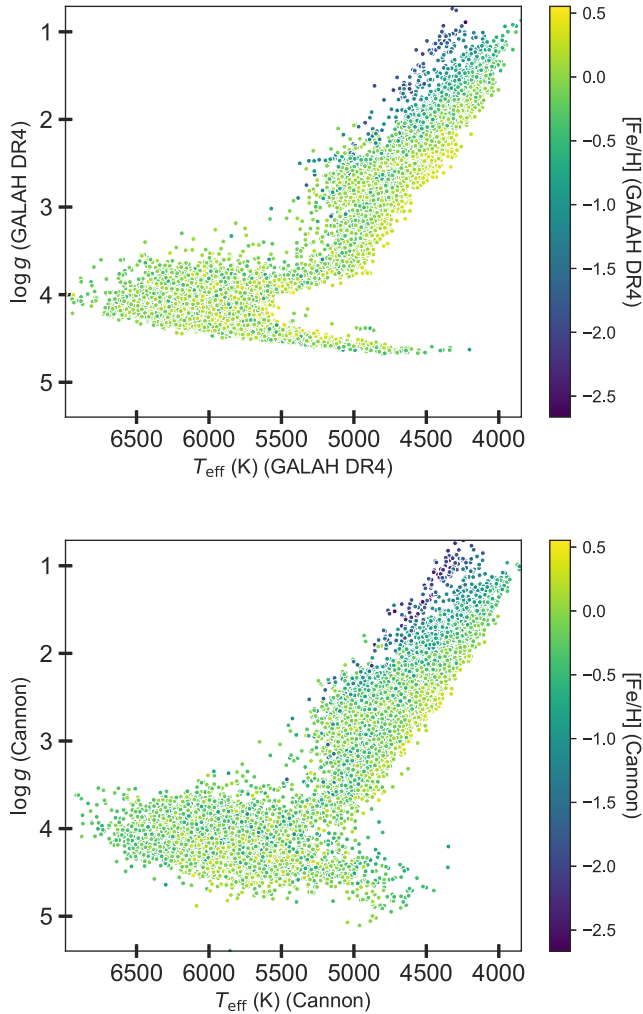

**Figure 2.** Comparison of Kiel diagrams (between *The Cannon* predictions and GALAH DR4 labels) for the training sample of 14 484 stars common to both the GALAH DR4 and *Gaia* DR3 RVS data sets. Top panel: Kiel diagram for the training sample, using $T_{\text{eff}}$, $\log g$, and [Fe/H] from GALAH DR4. Bottom panel: Kiel diagram for the training sample using *The Cannon* predictions ($T_{\text{eff}}$, $\log g$, and [Fe/H]), i.e. these are the stellar labels estimated from the low-resolution *Gaia* RVS spectra using the high-resolution GALAH stellar labels as model parameters in *The Cannon*.

the limitations when dealing with hotter stars. Finally, we require the selected sample of stars to have small uncertainties associated with GALAH DR4 stellar labels to ensure data quality and reliability (refer to Table 1 for the adopted quality cuts for different stellar labels from GALAH). The top panel of Fig. 2 is the Kiel diagram

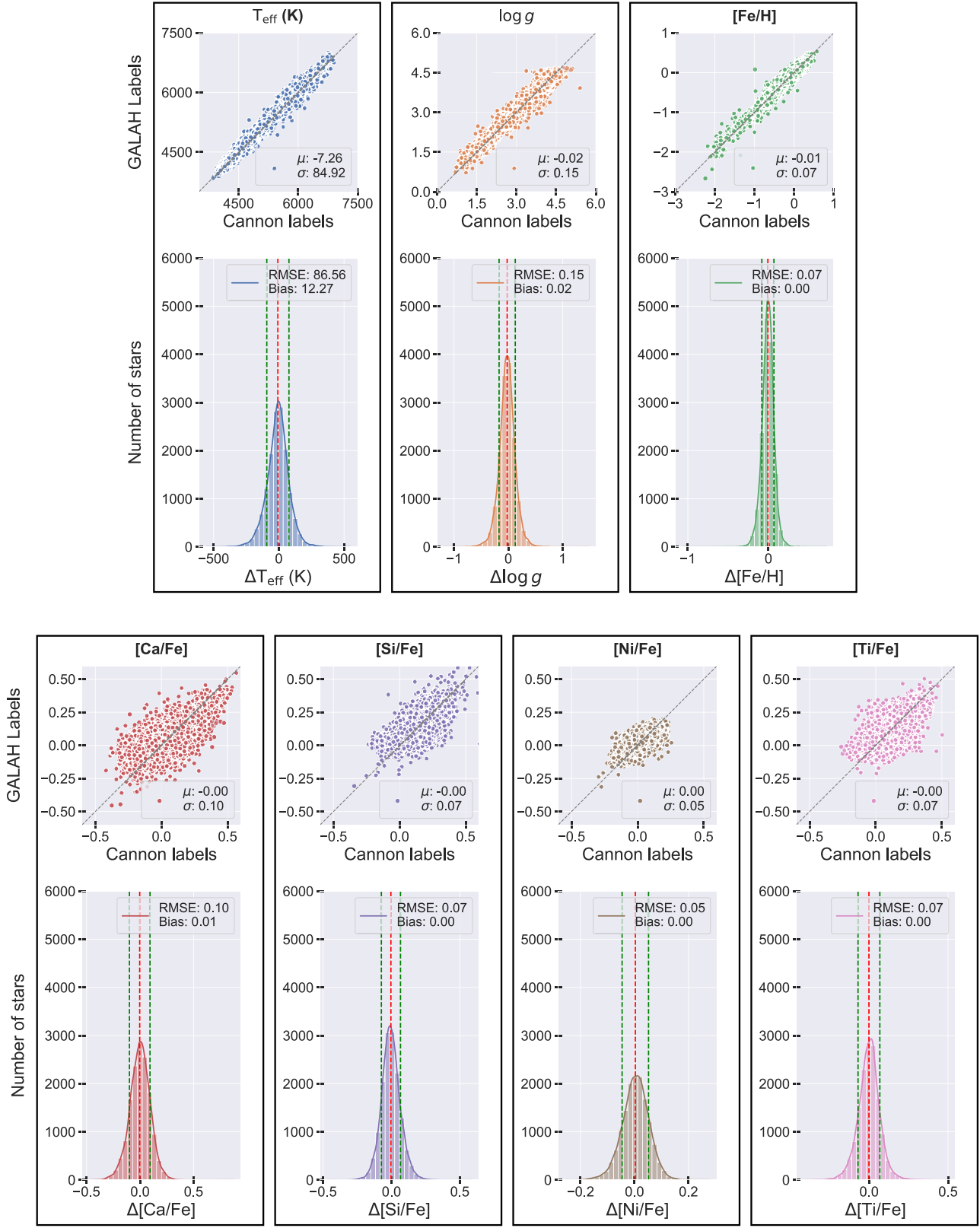

**Figure 3.** One-to-one relationships between $T_{\rm eff}$, $\log g$, [Fe/H], [Ca/Fe], [Si/Fe], [Ni/Fe], and [Ti/Fe] are illustrated in each rectangular box (outlined in black). Each box contains two plots for the stellar labels. The upper plot inside each box displays *The Cannon* predictions (*x*-axis) versus the corresponding stellar labels from GALAH DR4 (*y*-axis). $\mu$ and $\sigma$ represent the mean and standard deviation of the residual values (difference between *The Cannon* predictions and GALAH DR4), respectively. The lower plot within each box shows the distribution of these residuals across the total sample, including the RMSE and bias values obtained through 12-fold cross-validation (refer to Section 3.4). The residuals are centred with a prominent peak near zero, indicating the high accuracy of *The Cannon*'s stellar label predictions across all stellar parameters and abundances.

**Table 3.** Means ($\mu$) and standard deviations ($\sigma$) of the residual values ($T_{eff(Cannon)} - T_{eff(GALAH-DR4)}$ for stellar labels of the training sample).

| Stellar label | All | | Giants | | Dwarfs | |
|---|---|---|---|---|---|---|
| (*Cannon*−GALAH) | $\mu$ | $\sigma$ | $\mu$ | $\sigma$ | $\mu$ | $\sigma$ |
| $T_{eff}$ (K) | −7.26 | 84.92 | −5.74 | 69.95 | −10.47 | 109.95 |
| log g (dex) | −0.02 | 0.15 | −0.02 | 0.14 | −0.02 | 0.17 |
| $v \sin i$ (km $s^{-1}$) | −0.22 | 3.16 | −0.37 | 3.22 | 0.10 | 3.00 |
| [Fe/H] (dex) | −0.01 | 0.07 | 0.00 | 0.07 | −0.01 | 0.08 |
| [Ca/Fe] (dex) | 0.00 | 0.10 | 0.01 | 0.10 | 0.01 | 0.09 |
| [Si/Fe] (dex) | 0.00 | 0.07 | 0.01 | 0.07 | 0.00 | 0.07 |
| [Ni/Fe] (dex) | 0.00 | 0.05 | 0.01 | 0.05 | 0.00 | 0.05 |
| [Ti/Fe] (dex) | 0.00 | 0.07 | 0.00 | 0.07 | 0.00 | 0.07 |

*Notes.* Columns 2 and 3 list the means and standard deviations of the stellar labels derived from the entire sample. Similarly, columns 4 and 5 provide the corresponding metrics for giants (with log g < 3.5 dex), while columns 6 and 7 show the same quantities for dwarfs (with log g > 3.5 dex).

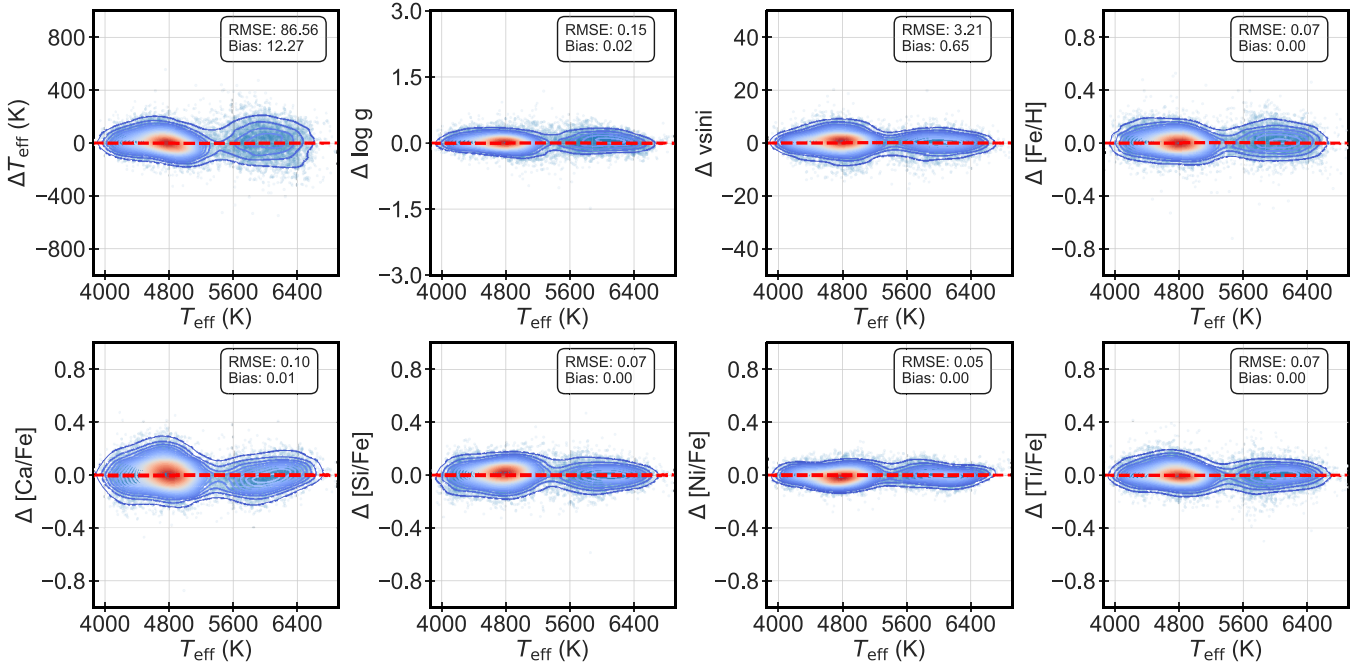

**Figure 4.** Residual plot for $T_{eff}$, log g, [Fe/H], $v \sin i$, [Fe/H], [Ca/Fe], [Si/Fe], [Ni/Fe], and [Ti/Fe], respectively. The *x*-axis indicates the stellar $T_{eff}$ labels from GALAH DR4, and the *y*-axis denotes the difference between *The Cannon* predictions and GALAH DR4 values for each of the stellar labels. The two separate blobs in each of the panels represent giants (at lower $T_{eff}$) and dwarfs (at higher $T_{eff}$). Residual values show similar scatter for both giants and dwarfs, with $\Delta T_{eff} \sim 84.92$ K, $\Delta$ log g $\sim 0.15$ dex, $\Delta$[Fe/H] $\sim 0.07$ dex, and other elemental abundances for Ca, Si, Ti, and Ni constrained within $\sim 0.07$ dex. A comparison of the statistics of the residuals between the giants and dwarfs is provided in Table 3.

including the corresponding RMSE and bias values. $T_{eff}$ shows a standard deviation of 84.92 K in residuals, and log g shows residual dispersion of 0.15 dex. We also observe small mean dispersion in the residual values for all the stellar abundances in [Fe/H], [Ca/Fe], [Si/Fe], [Ni/Fe], and [Ti/Fe] (provided in Table 3), which supports a good agreement between the stellar labels in the GALAH and *The Cannon* predictions.

Fig. 4 shows the trends in the residual values in $T_{eff}$, log g, $v \sin i$, [Fe/H], [Ca/Fe], [Si/Fe], [Ni/Fe], and [Ti/Fe] with $T_{eff,GALAH DR4}$ on the *x*-axis and the difference between *The Cannon*-predicted stellar labels and the input GALAH DR4 stellar labels on the *y*-axis. Our analysis reveals minimal standard deviations in the residual values, indicating a close agreement between the predictions of *The Cannon* and GALAH DR4 data sets. Notably, the mean residual differences across all stellar abundances converge to zero, reflecting consistent predictions by *The Cannon* across the total range of $T_{eff,GALAH DR4}$. In Fig. 4, we can also see two distinct groups within the residual values in each of the stellar labels, where the lower $T_{eff}$ range mainly corresponds to giants, while the higher $T_{eff}$ group mostly encompasses dwarfs. The cluster of points for the giants appears denser, owing to the larger fraction of giants in our total training sample ($\approx 60$ per cent).

Table 3 provides a comprehensive summary of the results for the training set, presenting the means and standard deviations of all the estimated stellar labels derived from the entire sample, as well as these quantities separately for giants (with log g < 3.5 dex), and dwarfs (with log g > 3.5 dex). Our analysis indicates a similar spread in residual values across all stellar abundances for both giants and dwarfs, with a somewhat smaller scatter in $T_{\rm eff}$ for giants (69.95 K), which is close to the mean scatter for the total training sample (84.92 K), mainly due to the majority of giants in the training sample.

### 3.5 Application of *The Cannon* to *Gaia* RVS spectra

Next, with the training set finalized, we used the model to predict the values for the much larger test sample, consisting of 796 633 *Gaia* RVS targets. We provided the *The Cannon* model with spectral fluxes (*flux*) and the corresponding flux errors (*err_flux*) for these *Gaia* RVS targets, which predicted the values for the stellar labels $T_{\rm eff}$, $\log g$, $v \sin i$, [Fe/H], [Ca/Fe], [Si/Fe], [Ni/Fe], and [Ti/Fe]. Plots of the reduced $\chi^2$ distribution for the model spectra and a few examples of model-predicted spectra compared to the corresponding observed RVS spectra are presented in Appendices A3 and A4, respectively.

To ensure that model predictions are within the confines of the training set, it is imperative to flag *The Cannon* estimates that extend beyond the bounds of the training set labels. This precaution is necessary, as the reliability of the predicted values from *The Cannon* decreases – in some cases, dramatically – when extrapolated to regions of parameter space well outside those covered by the training set. Therefore, for each of the targets in the test sample, we computed the distance, $D$ (Ho et al. 2017; Buder et al. 2018), between the corresponding *Cannon* estimates, $l_{\rm Cannon}$, and the training set labels, $l_{\rm GALAH}$, utilizing the following equation:

$$D = \sum_l \sum_{l_{\rm GALAH}} \frac{(l_{\rm Cannon} - l_{\rm GALAH})^2}{K_l^2}. \quad (8)$$

In this context, $K_l$ represents the uncertainties associated with each label, for which we utilized the RMSE values computed for the training set (as listed in Table 2). Employing the parameters $T_{\rm eff}$, $\log g$, $v \sin i$, [Fe/H], [Ca/Fe], [Si/Fe], [Ni/Fe], and [Ti/Fe] as the label space $l$, the average $D$ was calculated towards the closest 10 stars in the training set. We designated all *Cannon* estimates that exceeded 16 (a mean of 2$\sigma$ for the eight labels) as flag_cannon = 1, while those lesser than 16 were marked as flag_cannon = 0 in our final catalogue.

From our analysis, out of the 796 633 *Gaia* RVS targets, 610 823 stars in the test sample had flag_cannon = 0, indicating their alignment with the parameter space spanned by the label estimates derived from the GALAH training set. Among the stars with flag_cannon = 1, most of them have one or more of the following properties: $T_{\rm eff}$ > 7000 K; $T_{\rm eff}$ < 4500 K and ~2 < log g < 5 (i.e. cool dwarfs); high $v \sin i$; and elemental abundances outside the label space of the training sample. Fig. 5 shows the Kiel diagram for the 610 823 *Gaia* RVS stars (flag_cannon = 0), with the predicted stellar label log g on the y-axis and $T_{\rm eff}$ on the x-axis.

We compared our predicted labels $T_{\rm eff}$, log g, and [Fe/H] against those derived by Guiglion et al. (2024) in Fig. 6. Here, they employed a hybrid Convolutional Neural Network (CNN) that combined the *Gaia* DR3 RVS spectra, as well as *Gaia* DR3 photometry, parallaxes and XP spectroscopic coefficients to derive atmospheric parameters, [Fe/H], and composite $\alpha$-elemental abundances. We employed the recommended flags by setting flag_boundary = '00000000', and then compared the stellar labels with *The Cannon* estimates. In Fig. 6, the top panel shows the one-to-one comparisons between the *The Cannon*-estimated stellar labels for $T_{\rm eff}$, log g, and [Fe/H] and the corresponding values from Guiglion et al. (2024); the bottom panel shows the distribution of the residual values for the stellar labels with respect to $T_{\rm eff, CNN}$. We can see that, while there is good general agreement between the two studies, the correspondence worsens for stars at $T_{\rm eff}$ > 6200 K. For these stars, the $T_{\rm eff}$ measured by Guiglion et al. (2024) appears to plateau at around 6200 K. However our results appear to be robust up to nearly 7000 K, which is the upper limit of the GALAH $T_{\rm eff}$ parameter space.

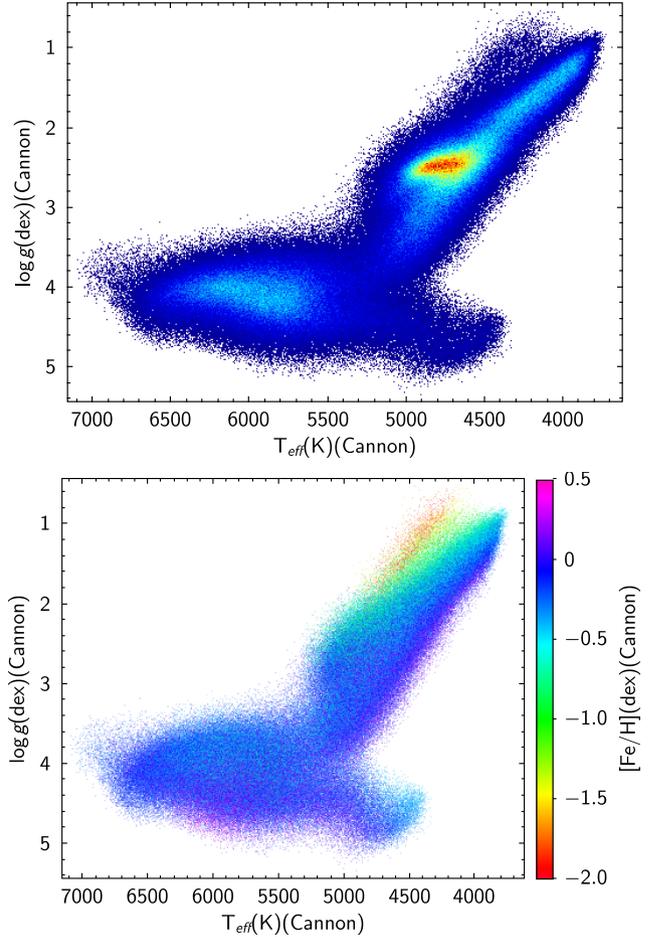

**Figure 5.** Kiel diagram for the test sample predictions (610 823 *Gaia* RVS targets with flag_cannon = 0). Top panel: the density map of *The Cannon* predictions for the test sample. Red clump stars are clearly visible between $T_{\rm eff}$ ~ 4500 − 5000 K and 2 < log g (dex) < 3. Bottom panel: the distribution of the estimated metallicity ([Fe/H]) for the test sample in the Kiel diagram.

We also compared our *Cannon* estimates for the stellar labels against Recio-Blanco et al. (2023), who used the General Stellar Parametriser-Spectroscopy (GSP-Spec) module to estimate the chemo-physical parameters from combined RVS spectra of single stars, without using any additional inputs from astrometric, photometric, or spectro-photometric BP/RP data. We used $T_{\rm eff}$ and the calibrated values of log g, [Fe/H], [Ca/Fe], [Si/Fe], [Ni/Fe], and [Ti/Fe] from Recio-Blanco et al. (2023). We have also used the recommended GSP-Spec flags (refer to table 2 from Recio-Blanco et al. 2023). Fig. 7 shows the density maps for the residual plots for the stellar labels versus the $T_{\rm eff}$ values from Recio-Blanco et al. (2023), along with the means and scatter of the residual values. We observe a trend in the scatter in the estimated $T_{\rm eff}$ between the two studies at higher temperatures, which is primarily because the training sample used for our model has a higher fraction of giants, with lower $T_{\rm eff}$ and log g (dex) < 3.5. For [Ca/Fe], we also observe a small increase in the residuals for stars with $T_{\rm eff} \gtrsim 5600$ K.

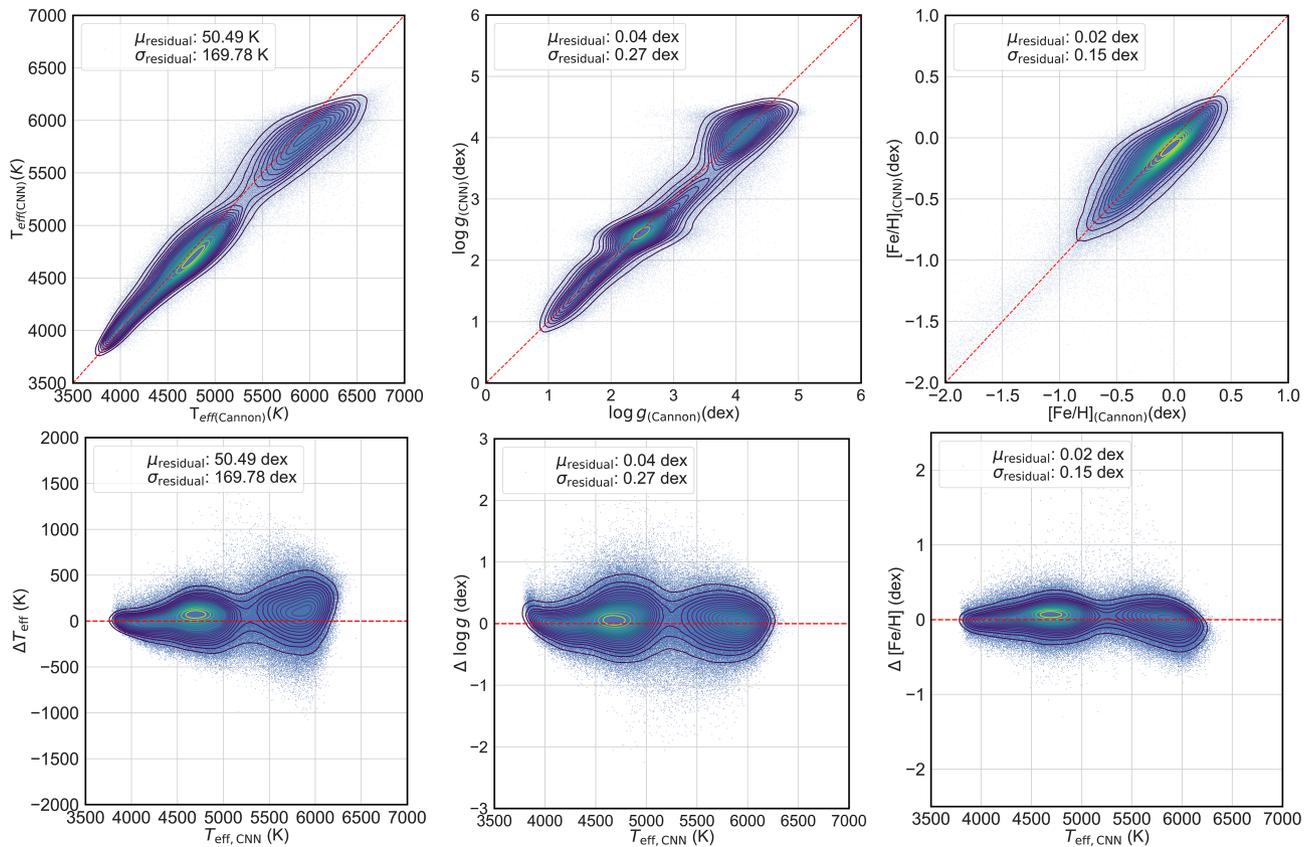

**Figure 6.** Top panels: Comparison of *The Cannon* estimated labels $T_{\rm eff}$, $\log g$, and [Fe/H] (*x*-axis) plotted against the predictions from Guiglion et al. (2024) (*y*-axis), who used a hybrid CNN to derive their stellar labels. Bottom panels: Density maps for the residual values for *The Cannon* predictions as compared to predictions from Guiglion et al. (2024); and $\mu$ and $\sigma$ represent the corresponding mean and standard deviation of the residual values of the stellar labels, respectively.

### 3.6 Error estimation

The covariance errors generated by *The Cannon* do not account for the overall systematic uncertainties, and therefore require a more statistical approach to error estimation for our results. To address this, for every spectrum in our test sample, we redrew each flux value from a normal distribution with a mean of its value (from *Gaia* RVS) and a 1$\sigma$ standard deviation corresponding to the error. We generated 10 such copies of each of the RVS stellar spectra and used *The Cannon* to derive 10 separate realizations of stellar labels for these stars. Then, the standard deviations of these realizations were determined for each star, which provided the S/N-dependent uncertainties of the predicted stellar labels. We carried this out on a random sample of 10 000 *Gaia* RVS targets, generating a total of 100 000 spectra from the normal distribution of the spectral flux in the process. Figs A1 and A2 show the exponential fits to the variation of the standard deviation in the stellar abundances ([Fe/H], [Ca/Fe], [Si/Fe], [Ni/Fe], and [Ti/Fe]), and the stellar parameters ($T_{\rm eff}$ and $\log g$) versus the *Gaia* S/N of these 10 000 RVS stellar spectra. As might be expected, we see a steep decline in the standard deviations for the stellar labels with increasing *Gaia* S/N. We observe a similar exponential decrease for the standard deviations of all stellar abundances [X/Fe] (where X = Ca, Si, Ni, and Ti) with increasing S/N, although the rate of decline for [Fe/H] is nearly twice that for other elements, suggesting that [Fe/H] is both very sensitive to S/N and accurately measurable (with small standard deviations) even at relatively low S/N.

## 4 VALIDATION OF THE TEST SAMPLE

With *The Cannon* estimates of stellar labels for the 610 823 *Gaia* RVS targets in hand, we set out to validate those stellar labels by comparing our results for members of a sample of globular and open clusters from the literature. Specifically, we focused on four globular clusters from the *Gaia* EDR3 globular cluster catalogue of Vasiliev & Baumgardt (2021) and six open clusters from the open cluster member catalogue provided by Spina et al. (2021) that overlap with our RVS test sample and have been previously studied. We focused our analysis primarily on the metallicities of these clusters.

### 4.1 Globular clusters

We selected the globular clusters NGC 104 (47 Tucanae), NGC 3201, NGC 6121 (M4), and NGC 6752 following a cross-match between the *Gaia* EDR3 globular cluster catalogue and our test sample, considering only those stars with cluster membership probabilities `memberprob > 0.9` (Vasiliev & Baumgardt 2021). We also included only the targets with `flag_cannon = 0`, and limited our selection to clusters with a cross-match sample size of at least 10 stars per cluster for our analysis.

Table 4 provides a summary of the predicted mean [Fe/H] values for the globular clusters alongside their corresponding literature values, and Fig. 8 shows the metallicities for the four globular clusters estimated by *The Cannon*, as well as from multiple literature references. Notably, for NGC 104, the mean predicted [Fe/H] is more metal-rich ([Fe/H] = $-0.63 \pm 0.02$ dex) in contrast to the

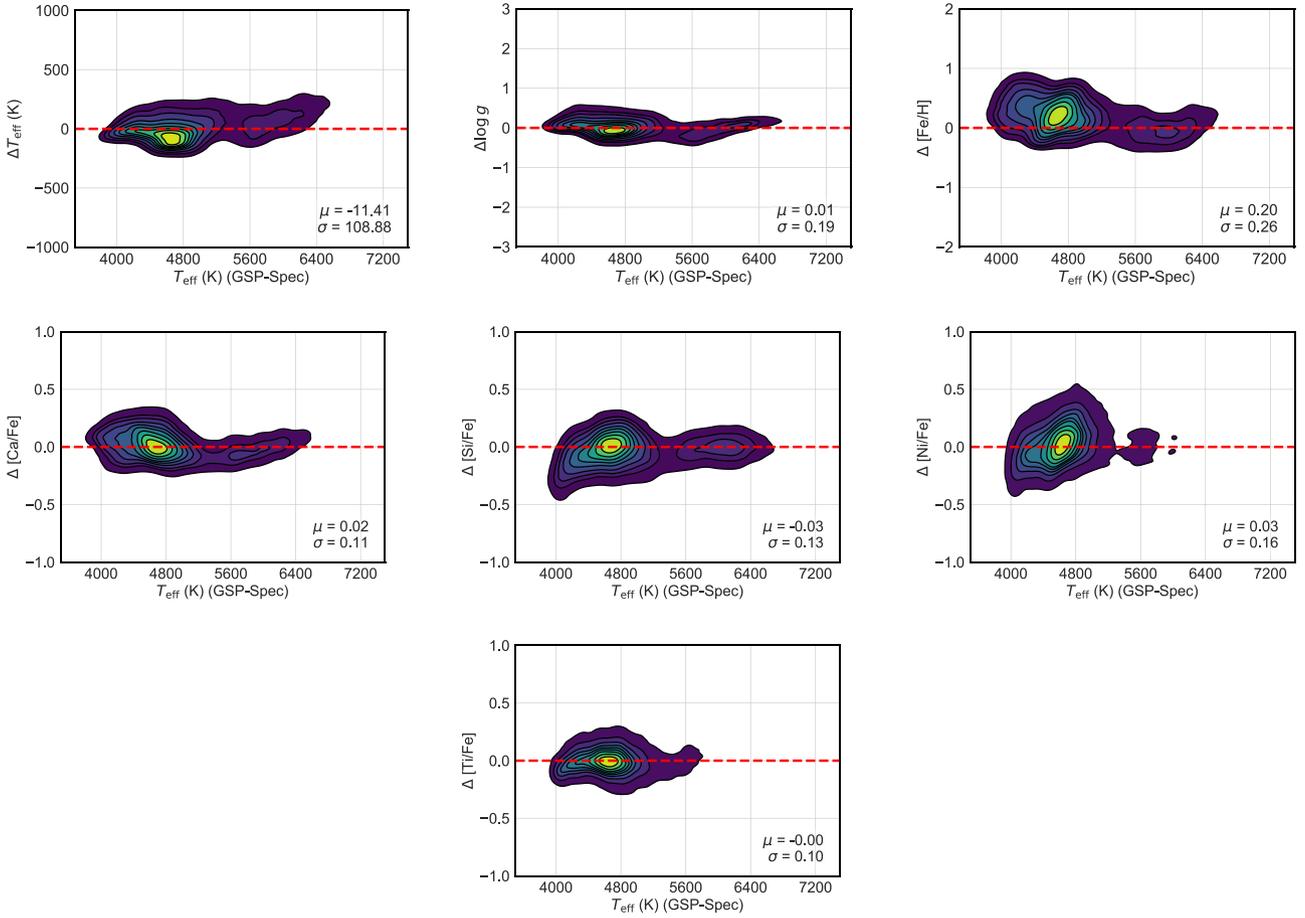

**Figure 7.** Density maps for the residual values for *The Cannon* predictions of $T_{\rm eff}$, log $g$, [Fe/H], [Ca/Fe], [Si/Fe], [Ni/Fe], and [Ti/Fe] for the *Gaia*–RVS spectra compared with the results from Recio-Blanco et al. (2023), who used the GSP-Spec module to estimate stellar labels. The *x*-axis indicates $T_{\rm eff}$ from Recio-Blanco et al. (2023), and the *y*-axis shows the residuals (*The Cannon* predictions − GSP-Spec results) for each of the stellar labels. $\mu$ and $\sigma$ represent the mean and standard deviation of the residual values of the labels, respectively.

**Table 4.** *The Cannon*-predicted metallicities for four globular clusters from the cross-matched sample of our test targets and the *Gaia* EDR3 globular cluster catalogue (Vasiliev & Baumgardt 2021).

| Cluster | [Fe/H]$_{\rm Cannon}$ (dex) | [Fe/H]$_{\rm Cannon:S/N>30}$ (dex) | [Fe/H]$_{\rm literature}$ (dex) | Source |
| --- | --- | --- | --- | --- |
| NGC 104 | $-0.63 \pm 0.02$ (82) | $-0.66 \pm 0.05$ (52) | $-0.76 \pm 0.02$ | 1 |
| (47 Tucanae) | | | $-0.70 \pm 0.03$ | 2 |
| NGC 6121 | $-0.93 \pm 0.03$ (24) | $-0.95 \pm 0.04$ (21) | $-1.18 \pm 0.02$ | 1 |
| (M4) | | | $-1.19 \pm 0.03$ | 2 |
| | | | $-1.05$ | 3, 4 |
| | | | $-1.07 \pm 0.01$ | 5 |
| NGC 3201 | $-1.27 \pm 0.04$ (14) | $-1.29 \pm 0.06$ (12) | $-1.51 \pm 0.01$ | 1 |
| | | | $-1.23 \pm 0.05$ | 2 |
| | | | $-1.24$ | 3,4 |
| NGC 6752 | $-1.36 \pm 0.09$ (19) | $-1.32 \pm 0.12$ (14) | $-1.55 \pm 0.01$ | 1 |
| | | | $-1.42 \pm 0.02$ | 2 |
| | | | $-1.24$ | 3, 4 |

*Notes.* The numbers in the brackets in columns 2 and 3 indicate the number of stars used for obtaining the mean metallicities. Column 5 contains the literature values for [Fe/H] for the clusters from multiple sources numbered as: 1: Carretta et al. (2009), 2: Carretta & Gratton (1997), 3: Marín-Franch et al. (2009), 4: Forbes & Bridges (2010), and 5: Marino et al. (2008).

mean reported metallicities of $-0.76 \pm 0.02$ dex (Carretta et al. 2009), and $-0.70 \pm 0.03$ dex (Carretta & Gratton 1997). A similar trend is also seen for NGC 6121, where *The Cannon* estimates for [Fe/H] are more metal rich than literature sources. NGC 3201 appears to be consistent with the broad range of metallicities quoted in the literature (Carretta & Gratton 1997; Marín-Franch et al. 2009; Forbes & Bridges 2010). Finally, the range of measured metallicities for NGC 6752 from the literature (Marín-Franch et al. 2009; Forbes & Bridges 2010) straddles *The Cannon* predictions. Overall, we find *The Cannon* estimates for the mean [Fe/H] of each cluster to fall within ∼ 0.05 − 0.10 dex scatter of the mean values from multiple literature references, thereby supporting the validation of the metallicities obtained.

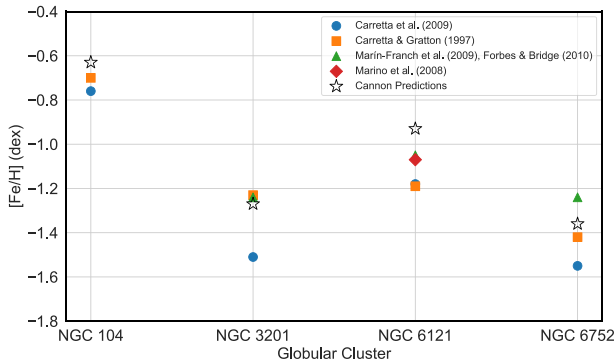

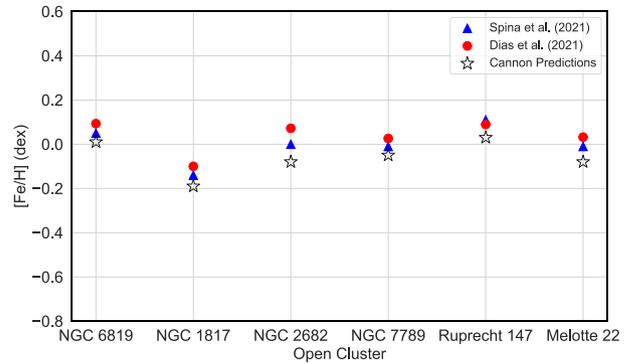

**Figure 8.** Mean metallicities for four globular clusters from the cross-matched sample of our test targets (*Gaia* S/N>30) and the *Gaia* EDR3 globular cluster catalogue by Vasiliev & Baumgardt (2021), with memberprob > 0.9 (probability of being a member of the cluster).

**Figure 9.** Mean metallicities for six open clusters from the cross-matched sample of our test targets (*Gaia* S/N>30) and the open cluster member catalogue provided by Spina et al. (2021) with memberprob > 0.9 (probability of being a member of the cluster).

### 4.2 Open clusters

We analysed the open clusters NGC 6819, NGC 1817, NGC 2682 (M67), NGC 7789, Ruprecht 147, and Melotte 22 following a cross-match between the open cluster member catalogue provided by Spina et al. (2021) and our test sample, and satisfying membership probability memberprob > 0.9. Similar to the analysis in the globular clusters, we only included targets with flag_cannon = 0, and selected clusters having a cross-match sample of at least 10 stars.

Table 5 shows the predicted mean metallicities for the open clusters alongside their respective literature values. An extra column has been included in the table, displaying the predicted [Fe/H] values exclusively for those targets with *Gaia* RVS S/N > 30. We note that the predicted mean metallicities of Ruprecht 147, NGC 7789, Melotte 22, and NGC 1817 are all lower (more metal-poor) compared to those reported in Spina et al. (2021). Melotte 22 yields the most metal-poor mean metallicity (−0.08 ± 0.02 dex) relative to the value from (Spina et al. 2021) (−0.01 ± 0.05 dex), with an offset of nearly

0.07 dex. Additionally, there is a noticeable reduction in the spread of predicted metallicities relative to the results for globular clusters (see Section 4.1), indicating improved precision in our predictive models; we attribute this at least in part to the fact that stars close to solar metallicity far outnumber metal-poor stars in our training sample. We also compared the metallicities of the six open clusters with Dias et al. (2021). Fig. 9 plots the metallicities predicted using *The Cannon*, as well those from Spina et al. (2021) and Dias et al. (2021).

## 5 DETECTION OF THE BIMODAL DISTRIBUTION IN [Ti/Fe]–[Fe/H] IN *Gaia* RVS DATA

Observations of the stellar populations of the Milky Way yield vital insights into the Galaxy's chemical composition and kinematic properties, which are essential for a comprehensive understanding of its formation and evolutionary history. One notable discovery was the identification of a thick disc component (Yoshii 1982; Gilmore & Reid 1983), which was initially recognized based on the vertical density distribution of stars in the Milky Way. Both Yoshii (1982) and Gilmore & Reid (1983) found that the stellar density could not be adequately described by a single exponential component, leading to the proposal of a second, thicker component. Although these

**Table 5.** *The Cannon*-predicted metallicities for the six open clusters selected from the cross-matched sample of our test targets and the open cluster member catalogue (Spina et al. 2021).

| Cluster | [Fe/H]$_{Cannon}$ (dex) | [Fe/H]$_{Cannon:S/N>30}$ (dex) | [Fe/H]$_{literature}$ (dex) | Source |
|---|---|---|---|---|
| NGC 6819 | 0.01 ± 0.03 (22) | 0.05 ± 0.03 (6) | 0.05 ± 0.03 | 1 |
| | | | 0.093 ± 0.006 | 2 |
| Ruprecht 147 | 0.03 ± 0.02 (42) | 0.05 ± 0.02 (33) | 0.11 ± 0.04 | 1 |
| | | | 0.089 ± 0.053 | 2 |
| NGC 7789 | −0.05 ± 0.02 (35) | −0.07 ± 0.02 (17) | −0.01 ± 0.02 | 1 |
| | | | 0.026 ± 0.028 | 2 |
| NGC 2682 / M67 | −0.08 ± 0.02 (46) | 0.01 ± 0.01 (15) | 0.00 ± 0.05 | 1 |
| | | | 0.072 ± 0.052 | 2 |
| Melotte 22 | −0.08 ± 0.02 (25) | −0.08 ± 0.02 (25) | −0.01 ± 0.05 | 1 |
| | | | 0.032 ± 0.029 | 2 |
| NGC 1817 | −0.19 ± 0.03 (21) | −0.22 ± 0.03 (13) | −0.14 ± 0.09 | 1 |
| | | | −0.1 ± 0.019 | 2 |

*Notes.* The numbers in brackets next to the metallicities in columns 2 and 3 indicate the number of stars used for obtaining the mean metallicities in each case. Column 5 contains the literature values for [Fe/H] for the clusters from multiple sources numbered as: 1: Spina et al. (2021) and 2: Dias et al. (2021).

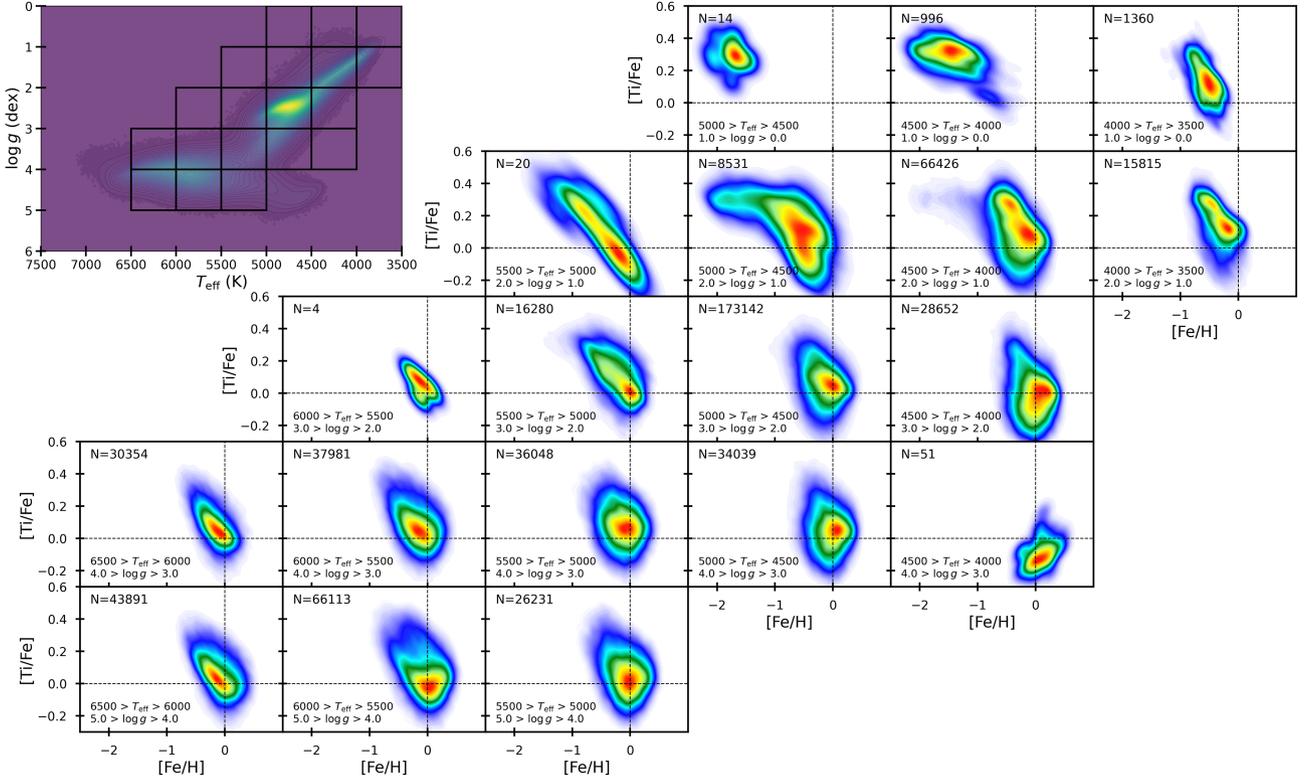

**Figure 10.** [Ti/Fe] versus [Fe/H] for a sample of 610 823 stars from the *Gaia* RVS survey satisfying `flag_cannon = 0` (see Section 3.5). In our work, we use [Ti/Fe] as the closest representative measure of [α/Fe]. The displayed data include stars that fall within the limits of the stellar label space of the training set. The panels are organized by effective temperature $T_{\rm eff}$ in increments of 500 K and surface gravity $\log g$ in intervals of 1 dex. $N$ represents the number of stars present inside each of the boxes in the panel. The Kiel diagram in the top-left corner illustrates the number density distribution of the 610 823 stars, with *The Cannon* estimates of $T_{\rm eff}$ (K) on the $x$-axis and $\log g$ (dex) on the $y$-axis. After fig. 14 of Guiglion et al. (2024).

early studies did not explicitly consider kinematics, later research demonstrated that the thick disc is distinct in both kinematics – exhibiting higher velocity dispersion and a different rotational lag (e.g. Schönrich & Binney 2009) – and chemical composition. The thick disc is characterized by a higher abundance of α-elements ([α/Fe]) at a given [Fe/H] compared to the thin disc (e.g. Fuhrmann 1998; Bensby et al. 2014; Hayden et al. 2015; Buder et al. 2019). This chemical distinction has led to the thick and thin discs often being referred to as the high- and low-α discs, respectively.

The origins of this dichotomy in the stellar populations of the Milky Way's disc have been and continue to be the subject of extensive debate within the astronomical community (e.g. Chiappini, Matteucci & Gratton 1997; Schönrich & Binney 2009; Minchev, Chiappini & Martig 2013; Noguchi 2018; Palla et al. 2020; Khoperskov et al. 2021). While it was not our intention with this work to weigh in on the nature of the high- and low-α discs, we did investigate whether our data set could potentially be applied to this and other problems in Galactic archaeology.

Using our newly derived abundances from the *Gaia* RVS sample, we examined the relationship between the elemental abundance ratio [α/Fe] and metallicity [Fe/H] within our observed sample of 610 823 stars to see if we could recover the expected bimodal distribution between high- and low-α discs. Guiglion et al. (2024) detected this [α/Fe] dichotomy in their analysis of *Gaia* RVS data. For our analysis, we used [Ti/Fe] as the best representative measure of [α/Fe] as its scatter is the smallest amongst the α-elements predicted using *The Cannon*. Fig. 10 shows the density map of the distribution of *The Cannon* predictions for [α/Fe] and [Fe/H] across different sections of the Kiel diagram for the 610 823 stars. We observe the bimodality in [Ti/Fe] most clearly in the region $4500 > T_{\rm eff}$ (K) $> 4000$ and $2 > \log g$ (dex) $> 1$, similar to Bensby et al. (2014), Hayden et al. (2015), Queiroz et al. (2020), and Guiglion et al. (2024).

Fig. 11(a) illustrates the distribution of [Ti/Fe] as a function of [Fe/H] for stars with $T_{\rm eff} < 5500$ K and $1 < \log g$ (dex) $< 3.8$. These criteria were selected to focus only on a sample of giant stars, following the example of Hayden et al. (2015). As expected, the data reveal distinct trends for the thick and thin disc stars in their [Ti/Fe] abundances at lower [Fe/H] values. Specifically, thick disc stars display higher [Ti/Fe] $> 0.2$ dex at lower [Fe/H] ($\lesssim -0.5$ dex), compared to thin disc stars. At approximately [Fe/H] $\sim -0.5$ dex, thick disc stars exhibit a notable decrease in their [Ti/Fe] values with increasing [Fe/H], a feature commonly referred to as the metallicity 'knee' (Hayden et al. 2015; Queiroz et al. 2020), eventually decreasing closer to the thin disc [Ti/Fe] abundances in the region of [Fe/H] $\sim 0.2$ dex.

To further investigate the different characteristic distributions of high- and low-[Ti/Fe] stars within the sample used for Fig. 11(a), we also plotted the histogram of the [Ti/Fe] distribution with bin sizes of 0.2 dex in [Fe/H], spanning a range of $-1.0$ to 0.2 dex in metallicity (see Fig. 11b). In this figure, two Gaussian functions are fitted to the [Ti/Fe] histogram in each metallicity bin (using Gaussian Mixture Model) to determine the peaks of the [Ti/Fe] distributions for the two stellar populations in our sample. For metal-poor stars ($-1.0 < $ [Fe/H] $< -0.8$ dex), the high-[Ti/Fe] group, associated with the thick disc, has a [Ti/Fe] ratio of $0.28 \pm 0.07$ dex, which distinguishes it from the low-[Ti/Fe] group, corresponding to the thin disc (0.12 ± 0.08 dex). The mean [Ti/Fe] abundance for thick disc stars remains relatively constant as [Fe/H] increases from −1.0 to −0.4 dex, but then decreases sharply from 0.23 to 0.16 dex over the [Fe/H] range from −0.4 and −0.2 dex, approaching solar metallicity with a mean [Ti/Fe] of 0.06 dex. The steep decline in [Ti/Fe] at [Fe/H] ≈ −0.5 dex indicates the 'knee' in the metallicity distribution, as shown in Fig. 11(a). For the low-[Ti/Fe] stars, the mean [Ti/Fe] ratio is initially enhanced within the metallicity range of −1.0 to −0.8 dex but gradually approaches solar values as [Fe/H] increases from −0.8 to 0.2 dex, in good agreement with the results of high-resolution spectroscopic studies such as Bensby et al. (2014).

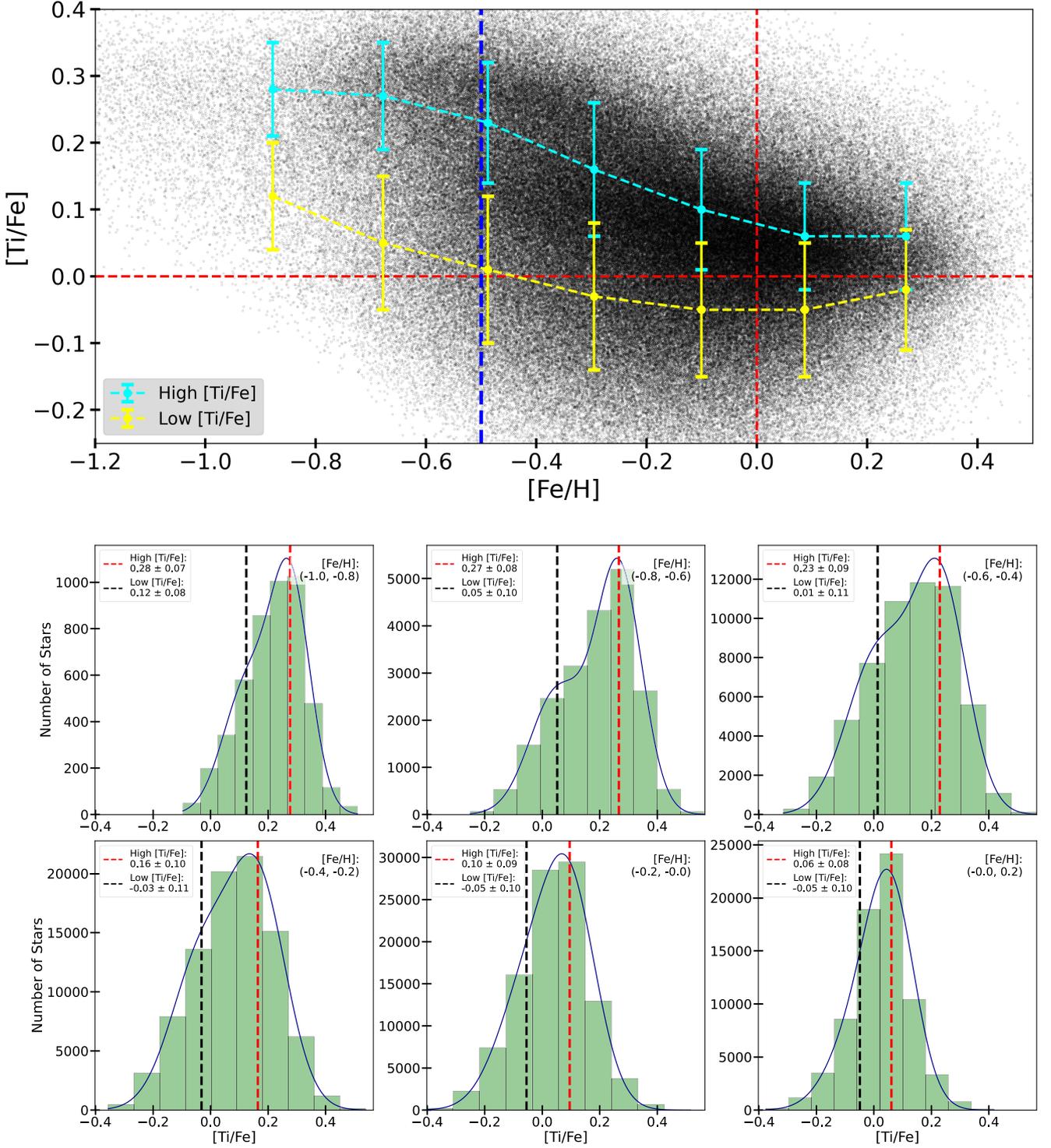

**Figure 11.** Bimodal distribution in [Ti/Fe] versus [Fe/H] observed in the Milky Way galaxy using *The Cannon*-predicted abundances of the stars in *Gaia* RVS. (a) [Ti/Fe] versus [Fe/H] for stars with `flag_cannon = 0` (refer to Section 3.5), $T_{\rm eff} < 5500$ K, and $1 < \log g$ (dex) $< 3.8$. The cyan and yellow dashed lines denote the mean [Ti/Fe] abundances for the high- and low-[Ti/Fe] disc stars respectively, which were obtained from the two Gaussian functions fitted to the sample (see below in part b). The dark-blue vertical dashed line at [Fe/H] $\sim -0.5$ dex denotes the metallicity 'knee', where the [Ti/Fe] abundances for the high-[Ti/Fe] stars begin to decrease with respect to [Fe/H]. This trend eventually converges close to the solar [Ti/Fe] ratio at [Fe/H] $\sim 0.2$ dex. The low-[Ti/Fe] stars show a characteristic upward concave shape (see, e.g. fig. 22 of Bensby et al. 2014). (b) Histogram distribution of [Ti/Fe] for giant stars (see part a). Each panel corresponds to stars within a metallicity bin of 0.2 dex, ranging from $-1.0$ to 0.2 dex (indicated in the top-right corner in each panel). Two Gaussian distributions are fitted for the data using Gaussian Mixture Model (GMM), with the mean of the higher and lower [Ti/Fe] groups represented by the red and black dashed lines, respectively. The mean and standard deviation for each Gaussian distribution are annotated in each panel for both the high- and low-[Ti/Fe] groups.

## 6 CONCLUSIONS

We have demonstrated the capability of *The Cannon* – a data-driven machine-learning approach – to link stellar properties derived from a high-resolution survey like GALAH DR4 ($R \sim 28\,000$) to data from a medium-resolution survey such as *Gaia* RVS ($R \sim 11\,500$). For developing *The Cannon* model, a training sample of 14 484 stars was selected common to both the *Gaia* DR3 RVS and the GALAH DR4 catalogues, based on multiple criteria provided in Section 3.2. We also performed 12-fold cross-validation tests on the training sample to obtain the statistics of the training sample and used the RMSE values to obtain the label distance for all the test sample stars (see Section 3.5). Following this, the model accurately determined stellar parameters ($T_{\rm eff}$ and $\log g$) and elemental abundances across a multidimensional label space, including [Fe/H], [Ca/Fe], [Si/Fe], [Ni/Fe], and [Ti/Fe], with a precision approximately ranging from 0.02 to 0.15 dex for 796 633 stars across the total sample with S/N > 15. In addition, we determined realistic error estimates for each stellar parameter and elemental abundance as a function of RVS spectrum S/N, a key consideration for a data set with a large fraction of low S/N spectra. We have shown the capability of *The Cannon* for predicting stellar labels precisely across a large S/N range, which is of particular importance considering that nearly one-third of the total *Gaia* RVS sample has S/N between 15 and 25 pixel$^{-1}$ element. Out of these 796 633 stars, a total of 610 823 stars fall within the limits of the stellar labels of the training sample. For scientific applications, we recommend using these 610 823 stars in our catalogue that are characterized by 'flag_cannon = 0'.

We were able to validate our predictions for [Fe/H] with multiple globular clusters (NGC 104/47 Tucanae, NGC 3201, NGC 6121/M4, and NGC 6752) and open clusters (NGC 6819, NGC 1817, NGC 2682/M67, NGC 7789, Ruprecht 147, and Melotte 22), where we demonstrated agreement with literature values within a scatter of ~ 0.02−0.10 dex (see Section 4, and Tables 4 and 5).

Using our resulting abundance catalogue, we were able to recover the characteristic bimodal distribution of [Ti/Fe] against [Fe/H] in disc stars, demonstrating the presence of two distinct disc populations – the high [α/Fe] (thick) and low [α/Fe] (thin) discs (Section 5, and Figs 10 and 11a). This is one of the first studies to observe such a clear dichotomy in [Ti/Fe] using *Gaia* RVS spectra (cf. Guiglion et al. 2024).

Finally, the catalogue presented in this paper illustrates the potential for exploiting *Gaia* RVS data to probe structures and substructures within the Milky Way. In particular, when combined with *Gaia* proper motions, parallaxes and radial velocities, our newly derived chemical abundances – probing a range of nucleosynthetic processes – provide a full chemo-dynamical data set for characterizing both *in situ* and accreted Galactic stellar populations. With the *Gaia* RVS data set expected to exceed 200 million stars by the end of the *Gaia* mission, *The Cannon* will likely be able to predict stellar parameters and abundances from RVS data efficiently and effectively over an even larger label space. In addition, *The Cannon* has the potential to yield useful results from lower S/N data, thereby greatly expanding the number of stars for which parameters and abundances can be obtained. In combinations with more traditional model-based methods, which will also provide parameters and chemical abundances, the two approaches are complementary and will contribute significantly to the chemical tagging of the galaxy, yielding a treasure trove for Galactic archaeology.

## ACKNOWLEDGEMENTS

PBD is supported by the Australian Government International Research Training Program (iRTP) Scholarship. Parts of this research were supported by the Australian Research Council Centre of Excellence for All Sky Astrophysics in 3 Dimensions (ASTRO 3D), through project number CE170100013. DBZ, AMG, GFL, and SLM acknowledge the support of the Australian Research Council through project number DP220102254. SLM acknowledges funding from the UNSW Scientia Fellowship program. DBZ thanks G. Guiglion for helpful conversations in the early stages of this project.

This work made use of the Fourth Data Release of the GALAH survey (Buder et al. 2024). The GALAH survey is based on data acquired through the Australian Astronomical Observatory, under programs: A/2013B/13 (The GALAH pilot survey); A/2014A/25, A/2015A/19, A2017A/18 (The GALAH survey, Phase 1); A2018A/18 (Open clusters with HERMES); A2019A/1 (Hierarchical star formation in Ori OB1); A2019A/15 (The GALAH survey, Phase 2); A/2015B/19, A/2016A/22, A/2016B/10, A/2017B/16, A/2018B/15 (The HERMES-TESS program); and A/2015A/3, A/2015B/1, A/2015B/19, A/2016A/22, A/2016B/12, A/2017A/14 (The HERMES K2-follow-up program). We acknowledge the traditional owners of the land on which the Anglo-Australian Telescope stands, the Gamilaraay people, and pay our respects to elders past and present. This paper includes data that has been provided by AAO Data Central (https://datacentral.org.au/).

This work has also used data from the European Space Agency (ESA) mission *Gaia* (https://www.cosmos.esa.int/gaia), processed by the *Gaia* Data Processing and Analysis Consortium (DPAC, https://www.cosmos.esa.int/web/gaia/dpac/consortium). Funding for the DPAC has been provided by national institutions, in particular the institutions participating in the *Gaia* Multilateral Agreement.

In addition, this paper used the NASA Astrophysics Data System (ADS) bibliographic services, as well as the open-source PYTHON packages ASTROPY (http://www.astropy.org) (Astropy Collaboration 2013, 2018), NUMPY (Harris et al. 2020), PANDAS (McKinney 2010), MATPLOTLIB (Hunter 2007), and TOPCAT (Taylor 2005, 2011).

## DATA AVAILABILITY

The data underlying this study are available at CDS via anonymous ftp to cdsarc.u-strasbg.fr (130.79.128.5) or via https://cdsarc.cds.unistra.fr/viz-bin/cat/J/MNRAS/538/605.

# APPENDIX A: COMPARISON OF MODEL SPECTRA GENERATED BY *THE CANNON* TO THE OBSERVED RVS SPECTRA

Figs A1 and A2 show the distribution of the standard deviations in the stellar abundances ($\sigma_{\mathrm{abundance}}$) and stellar parameters ($\sigma_{\mathrm{parameters}}$), respectively, with increasing S/N of the *Gaia* RVS spectra for a random sample of 10 000 stars.

Fig. A3 shows the distribution of reduced $\chi^2$ values obtained by comparing *The Cannon*-predicted spectra with the corresponding *Gaia* RVS spectra for 796 633 *Gaia* RVS targets.

Fig. A4 shows four examples of stellar spectra, spanning a range of metallicities, with the predicted spectrum from *The Cannon* in red, and the reference spectra from *Gaia* RVS in blue.

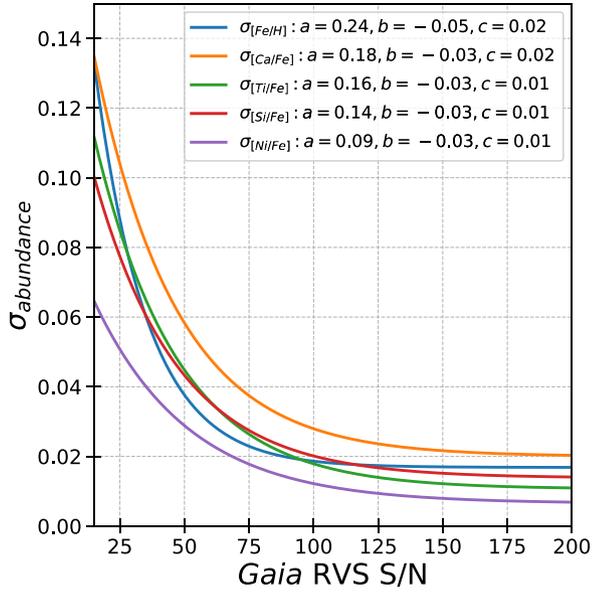

**Figure A1.** Exponential fits to the distribution of $\sigma_{\text{abundance}}$ with *Gaia* RVS S/N. The $\sigma_{\text{abundance}}$ for each target is obtained from the standard deviation of the predicted stellar abundances, derived from a Gaussian sample of flux values for each spectrum of a random selection of 10 000 targets from *Gaia* RVS. A steep decline in the standard deviation ($\sigma$ falling from $\sim 0.10$ to 0.05 dex) is evident at relatively low S/N $\approx 25$ for [Ca/Fe], [Ti/Fe], [Si/Fe], and [Ni/Fe]. The exponential fit applied is of the form $y = ae^{bx} + c$, where $y$ is the standard deviation of the abundance, $x$ denotes the *Gaia* RVS S/N, and the parameters $a$, $b$, and $c$ represent the initial amplitude/scaling factor, rate of exponential decay, and the baseline level that the standard deviation approaches at high S/N, respectively. A similar rate of decrease ($b$) of the standard deviation with increasing S/N is observed for all the abundances, with [Fe/H] showing the steepest decline ($b = -0.05$) from $\sim 0.15$ to 0.05 dex as it reaches S/N $\approx 30$.

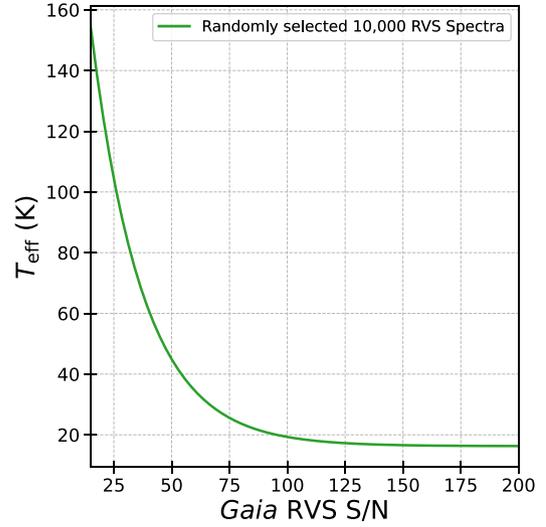

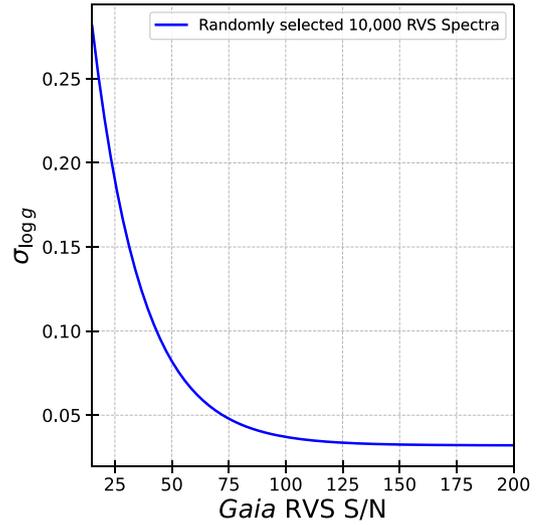

**Figure A2.** Similar to Fig. A1, the distribution of $\sigma_{\text{parameters}}$ ($T_{\text{eff}}$ and $\log g$) also shows a steep decline with increasing *Gaia* RVS S/N.

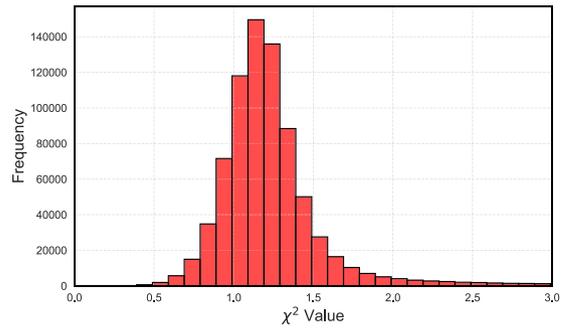

**Figure A3.** Distribution of reduced $\chi^2$ values between predicted spectra from *The Cannon* and the corresponding observed RVS spectra for the 796 633 *Gaia* RVS targets.

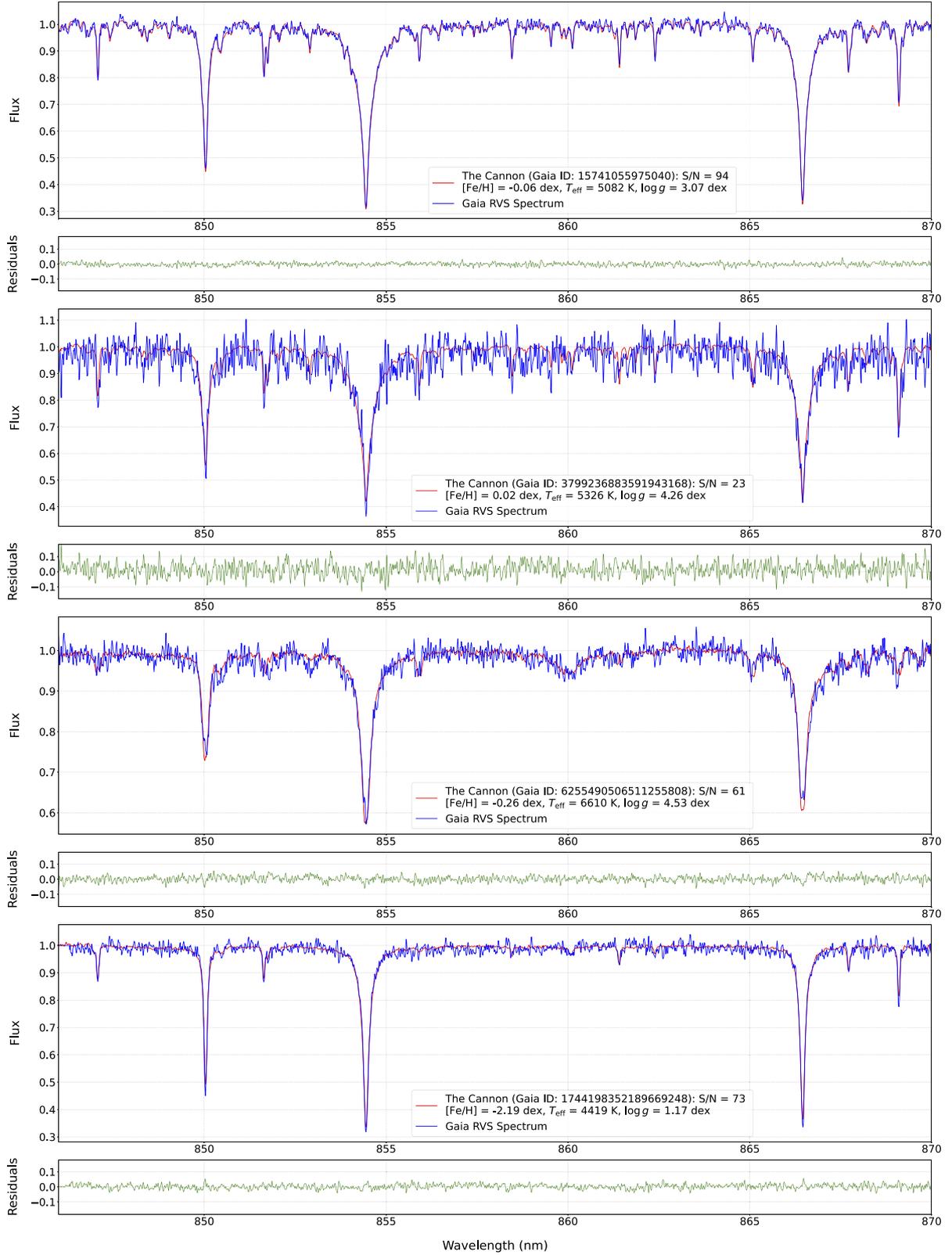

**Figure A4.** Comparison of the observed *Gaia* RVS stellar spectra (blue) against model-generated spectra by *The Cannon* (red), for four example stars with varying stellar labels (provided in each main panel) and the corresponding *Gaia* RVS S/N. The first two panels correspond to the flux and the residual values for the star *Gaia* DR3 15741055975040. Similarly, the next two panels are for the star *Gaia* DR3 3799236883591943168, followed by the two panels for *Gaia* DR3 6255490506511255808, and the last two panels are for the star *Gaia* 1744198352189669248.

This paper has been typeset from a TEX/LATEX file prepared by the author.